\documentstyle[twocolumn,aps,amssymb]{revtex}

\begin{document}
\title{Intrinsically frustrated superconducting array of SFS $\pi$-junctions}
\author{V. V. Ryazanov, V. A. Oboznov, A. V. Veretennikov, and A. Yu. Rusanov}
\address{Institute of Solid State Physics, Russian Academy of
Sciences, Chernogolovka, 142432, Russia}

\date{\today}
\maketitle
\begin{abstract}

We report a direct observation of the crossover of the superconductor-
ferromagnet-superconductor (SFS) junctions to the $\pi$-state which manifests 
itself in the half-period shift of the external magnetic field dependence of the 
transport critical current in the triangular SFS arrays. It is associated with 
the appearance of spontaneous supercurrents in the array at zero external field, 
with the ground state degenerated with respect to the two possible current flow 
directions. In conventional Josephson arrays this state can be observed only by 
imposing frustrating external field equal to a half-integer number of magnetic 
quanta per cell.
\end{abstract}

\section*{}
In recent years considerable attention has been given to realization of '$\pi$-contacts',
 i.e. weakly coupled junctions formed in superconducting structures 
which demonstrate $\pi$-shift of macroscopic phase difference in the ground 
state. Among such structures one should mention bicrystals \cite{ts4} and '$s$-$d$'
contacts \cite{vanH7} based on high-temperature 
superconductors with assumed 'nontrivial' ($d$-wave) symmetry of the order 
parameter, mesoscopic superconductor-normal metal-superconductor (SNS) junctions 
controlled by current along the N-layer \cite{Bas}, and finally, recently observed
superconductor-ferromagnet-superconductor (SFS) $\pi$-junctions \cite{LT,PRL}.
In the experiments  \cite{LT,PRL} the phase $\pi$-shift manifested itself in  
reentrant superconducting behavior of the critical supercurrent temperature 
dependence, $I_c(T)$, of the Nb-Cu/Ni-Nb Josephson SFS junctions. In this work we 
present direct evidence of the phase $\pi$-shift  which shows itself as a half-period
shift of the external magnetic field dependence of the transport critical 
current in a triangular SFS junction array at the point of the junction crossover  
from a '0'- to a '$\pi$'- state. This shift is associated with the appearance of 
spontaneous supercurrents in the array in case of zero external 
field, with the ground state degenerated with respect to the two possible 
current flow directions. In conventional Josephson arrays this state can be 
observed only by imposing a frustrating external field equal to a half-integer 
number of magnetic quanta per cell.	Self-frustrated superconducting networks 
with $\pi$-junctions are intended to be used for the realization of the superconducting 
quantum bit (qubit)\cite{Feig}. Originally, the suggested superconducting 
'phase' qubits \cite{Mooij}  were based on magnetic-frustrated 
superconducting networks. In this case it is necessary to imply half a quantum 
of the magnetic flux, $\Phi _0/2$, so as to create degenerated two-level coherent quantum system, 
requested for the realization of the qubit. This system is not isolated from 
interference with environment and is estimated to have shorter coherence time 
with respect to qubit using a $\pi$-junction. Another possible application of  
$\pi$-junctions is related to the development of superconducting digital 
'complementary' electronics where a $\pi$-junction is used as a superconducting 
phase inverter \cite{Terz}.

An origin of the $\pi$-state in a SFS junction is an oscillating and sign-reversal 
superconducting order parameter induced in the ferromagnet close to the SF-interface 
\cite{Andr,Kupr92,PRL}. It was shown in \cite{PRL} that in the case of weak 
exchange energy $E_{ex} \geq k_BT$, when the thermal and exchange 
energy make comparable contributions to the pair decay process, the general 
expression for the complex coherence length in the dirty ferromagnet is the 
following:

\begin{equation}
\xi_F = \left( \frac{\hbar D}{2(\pi k_BT + iE_{ex})}\right)^{1/2},
\label{5}
\end{equation}
where $D$ is the electron diffusion coefficient in the ferromagnet and $k_{B}$ 
is the Boltsman constant. The imaginary part 

\begin{equation}
\xi_{F2} = \sqrt{\frac{\hbar D}{((\pi k_BT)^2 + E_{ex}^2)^{1/2} - \pi k_BT } }
\label{6}
\end{equation}
defines the order parameter oscillation wavelength $2\pi\xi_{F2}$ that decreases 
with temperature. This allows observation of the crossover of the SFS junction to 
the $\pi$-state as temperature decreases down to $T=T_{cr}$, when the 
ferromagnetic interlayer thickness, $d_F$, becomes close to the half-wave of the 
oscillations, $\pi \xi _{F2}$.  It was detected in \cite{PRL} that the change of the 
state from the ordinary 0-phase state to the $\pi$-state leads to a zero-crossing 
of critical current temperature dependence, $I_c(T)$, or more truly to 
a sharp cusp with vanishing of the $I_c$ amplitude at $T=T_{cr}$, because 
only the absolute value of the critical current could be observed. Such reentrant temperature 
dependences of critical current were measured for $Nb-Cu/Ni-Nb$ Josephson 
sandwiches with ferromagnetic layers fabricated from Cu/Ni alloy with
concentrations from 52~at.\% Ni (with $T_{Curie}$ about 20 - 30~K)
to 57~at.\% Ni (where $T_{Curie}$ was above 100~K).

Several experimental phase-sensitive methods of direct detection of the $\pi$-state 
crossover were suggested. The original 
signature of the $\pi$-state is the appearance of a spontaneous magnetic flux equal to 
half of the flux quantum, $\Phi _0/2$, in a superconducting 
loop containing a $\pi$-contact. Josephson $\pi$-junction was first predicted in 
\cite{Bul} and it was shown there that the state with a 
spontaneous flux could be realized in such a one-contact interferometer only if 
$2\pi LI_c \gg \Phi_0$, where $L$ is the loop inductance 
and $I_c$ is the contact critical current. The other limit, $2\pi LI_c \ll 
\Phi_0$, more convenient in our case,  can be used to observe the shift of the external 
flux dependence of the transport critical current $I_m(\Phi)$ for a two-contact 
interferometer including Josephson $0$- and 
$\pi$-contacts. However fabrication of $0$- and $\pi$-junctions with close Josephson 
parameters in a two-junction SFS structure was a  rather difficult technical problem, and therefore 
we used a periodical array with three identical SFS junctions per cell. 
We discuss the more descriptive case of  the '$0$-$\pi$-interferometer' before consideration of 
the triangular array of the SFS junctions used in our experiment, because of close analogy of
the $0$-$\pi$-interferometer and two-cell $\pi$-junction triangular array behavior.
The presence of the phase difference of $\pi$ across the $\pi$-junction in the $0$-$\pi$-interferometer 
results in the circulating supercurrent close to the critical 
one and the extra phase shift of $\pi/2$ in both $0$- and $\pi$-junctions, hence 
the necessary additional $\pi$ phase difference appears in 
the superconducting loop. So the critical transport current $I_m$ through this 
interferometer ($0$-$\pi$-SQUID) is equal 
to zero in the absence of any external magnetic flux and reaches the maximum 
$I_m=2I_c$ when the external magnetic flux $\Phi$
is equal to $\Phi _0/2$. In the latter case the magnetic flux induces the additional phase shift 
of $2\pi(\Phi /\Phi _0)=\pi$ which compensates the spontaneous $\pi$-shift over 
the $\pi$-junction. Thus the $0-\pi$ interferometer 
initially is in a fully frustrated state.

The two-cell array, shown in Fig.1, is believed to be the most simple and plain structure to study the 
intrinsically frustrated networks. Details of the SFS junction fabrication are presented in \cite{PRL}.
All five $Nb-Cu_{0.46}Ni_{0.54}-Nb$ junctions in the array had areas 10$\times$10~$\mu$m$^2$, 
normal resistances 3$\times$ 10$^{-4}~\Omega$ and junction
critical currents $I_c$=0.3~$\mu$A at $T$=4.2~K. The $Cu_{0.46}Ni_{0.54}$ 
alloy with $T_{Curie}$ about $100$~K was used as a F-interlayer in the SFS contacts. Details of the
SFS junction fabrication were presented in \cite{PRL}. Because of the low array resistance 
and array critical current, $I_m$,  the current-voltage and $I_m(H)$ characteristics (H is magnetic field
applied normally to the array) were measured by
a SQUID picovoltmeter with a sensitivity of 10$^{-12}$~V in the temperature range of 1.2~K to 4.2~K.
The SFS junctions with the F-layer thickness of $d_F=19$~nm show the 
crossover to the $\pi$-state at $T_{cr}$=2.2~K. Above this temperature ($0$-
state) the $I_m(H)$ pattern is the same as the one predicted in 
\cite{Duzer,Likh} for a two-cell interferometer (see Fig.2a). Periodical maximal 
peaks were observed 
at external fields corresponding to an integer number of the flux quanta per 
cell, i.e. integer frustration parameters 
$f=\Phi / \Phi _0$. The small peaks, which one can see at half-integer frustration parameters,
correspond to quantization on the contour of 
the net structure which is twice the cell square. Below $T_{cr}$ (in the 
$\pi$-state) the $I_m(H)$ pattern was found 
to be shifted by exactly half a period (Fig.2b). Without considering the 
quantization process for the doubled cell square (that resulted 
in the appearance of the small peaks), the behaviour of this $\pi$-junction 
interferometer resembles that of the $0$-$\pi$-interferometer 
described above. At $f=0$ and other integer values of $f$ there is a current 
close to the junctions critical current, $I_c$, flowing 
in the outer double loop of the array. The current induces the extra phase shift 
$2\times \pi /2$ in each cell and compensates for the 
odd number of $\pi$-shifts in them. The maximal array transport 
supercurrent, $I_m$,  is close to zero at $H$=0 because the array is
initially in the spontaneous fully frustrated state. The external magnetic flux 
equal to half-integer quanta per cell produces
the necessary phase shift of $\pi$ in each cell in the absence of any circular 
currents in the structure, therefore $I_m$ 
reaches maxima only at half-integer frustration parameters. Due to the
quantization on the doubled cell contour of the net structure $I_m(H=0)$ is not zero in $\pi$-state
but equal to amount of small peak, as it seen in Fig.2b.
The $I_m(T)$ 
dependence at $H=0$, shown in Fig.3a, is similar 
to the $I_c(T)$ dependence for single junctions and demonstrates the sharp cusp 
at the temperature $T_{cr}$ of the SFS 
junctions crossover to the $\pi$-state. However, one should take into account 
that while the high-temperature branch corresponds to the 
temperature dependence of the doubled junction critical current, 
the low-temperature branch 
corresponds to the small peak amplitude dependence on temperature. Fig.3b shows 
the maximal peaks positions before 
and after the crossover and demonstrates how abrupt it is.

To check possible residual magnetic inductance effects at low temperatures we fabricated also
a reference  two-cell array with the $Nb-Cu_{0.46}Ni_{0.54}-Nb$ SFS junctions which do not exhibit
a crossover to $\pi$-state in full experimental temperature 'window' (from 1.2 to 4.2~K) due to smaller
ferromagnetic interlayer thickness of 18~nm. After a proper careful cooldown (described in \cite{PRL})
zero-field value was the maximal  for $I_m$ because of formation of a
small-scale magnetic domain structure in the F-layer with zero average net magnetization. The $I_c(H)$
pattern was not shifted in experimental temperature range, and it proves that the shift observed
for the SFS junction array with $d_F$=19~nm is not associated with residual magnetic inductance
changes. Thus at temperature lower than the temperature of the crossover of the SFS junctions in the 
triangular array to the $\pi$-state the self-frustrated state is observed at zero external magnetic 
field and it is similar to the state of a conventional array frustrated by imposing an external field 
equal to a half-integer number of magnetic quanta per cell.
\\

\noindent The authors would like to thank J. Aarts, M.~V. Feigel'man and A.~A. Golubov for the helpful 
discussion and N.~S. Stepakov for the technical assistance. 
The work was supported by the state program Mesoscopic Electron Systems, INTAS
grant N 97-1940, and partially by Swiss National Research  Foundation project 
NFSF N7SUPJ062253.00 and by a grant from the Netherlands Organization for 
Scientific Research (NWO).

\begin{figure}

\caption{Real (upper) and schematic (low) picture of the network of five SFS
 junctions $Nb-Cu_{0.46}Ni_{0.54}-Nb$ ($d_{F}=19$~nm), which was used in
the phase-sensitive experiment.}

\end{figure}

\begin{figure}
\caption{Magnetic field dependences of the critical transport current for the structure
depicted in Fig.1 at temperature above (a) and below (b) $T_{cr}$.}
\end{figure}

\begin{figure}
\caption{(a) Temperature dependence of the critical transport current for the
structure depicted in Fig.3 in the absence of magnetic field; (b) temperature
dependence (jump) of the position of the maximal peak on the curves 
$I_{m}(H)$, corresponding to the two limiting temperatures depicted in Fig.2.}
\end{figure}

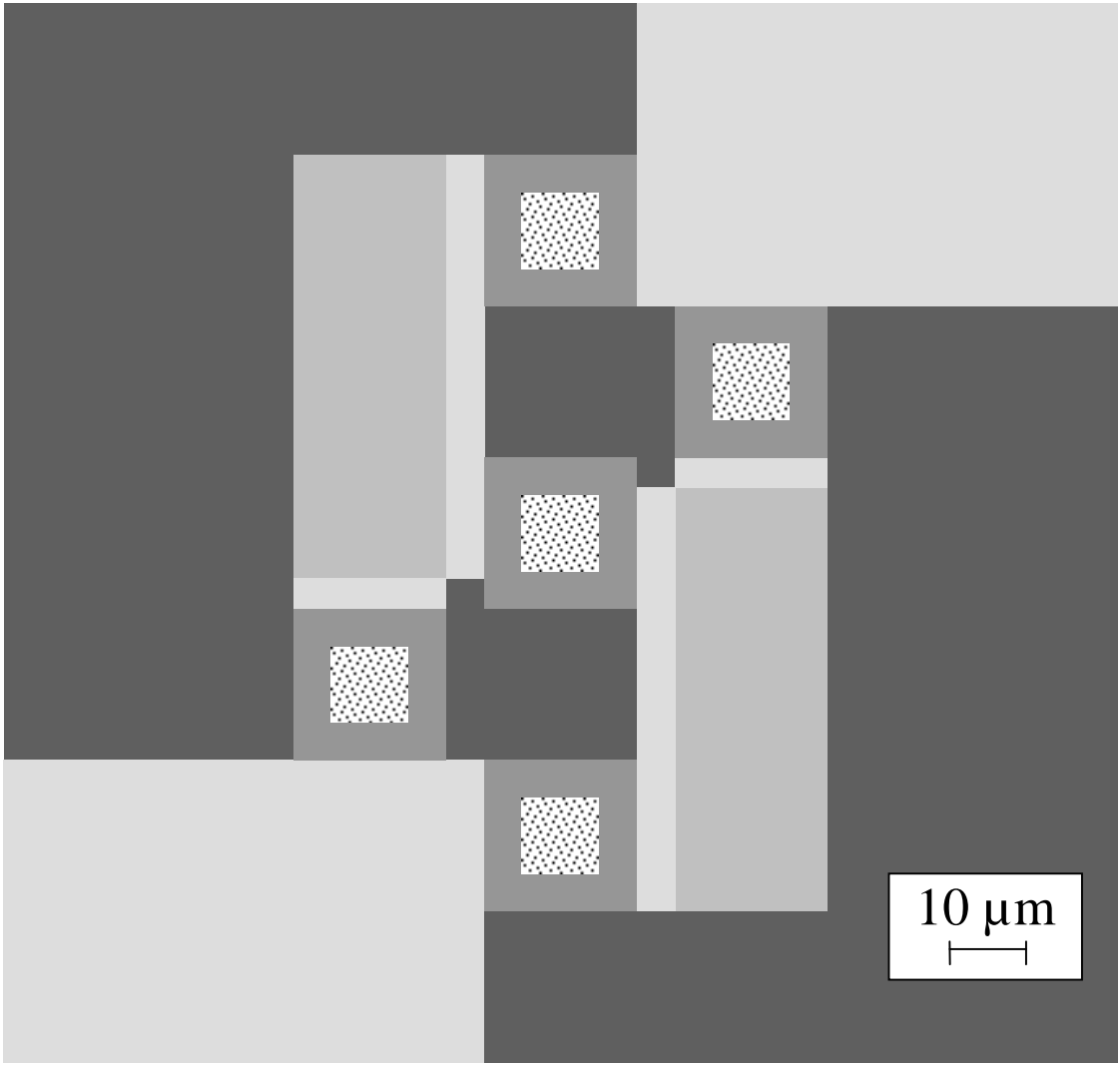
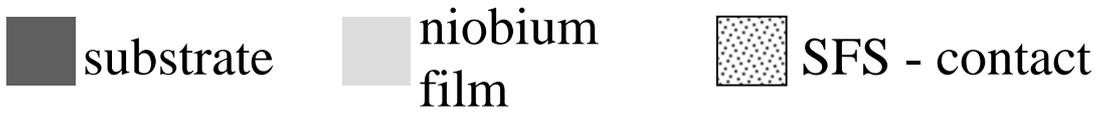
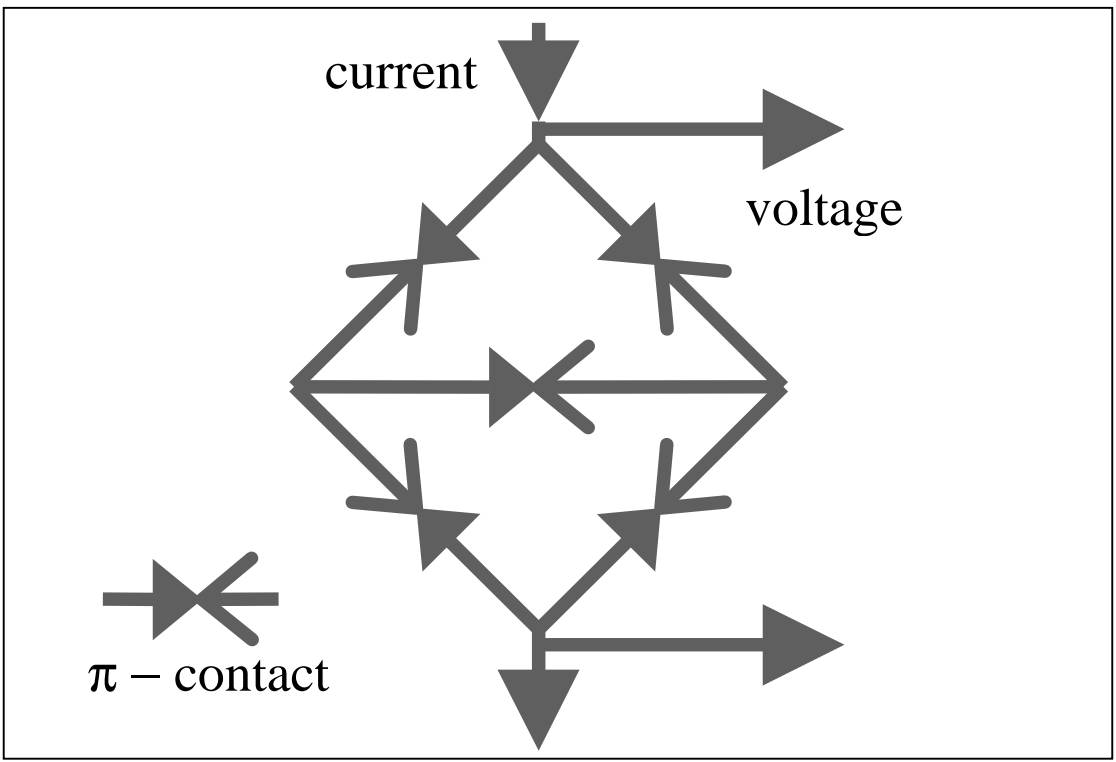
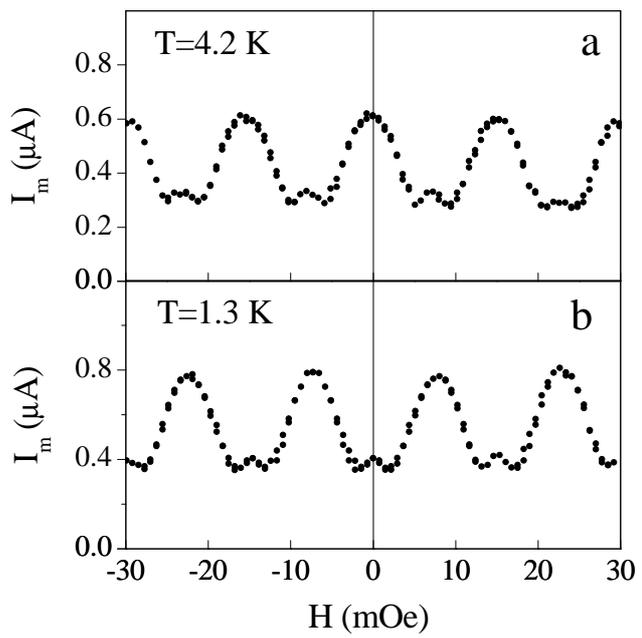
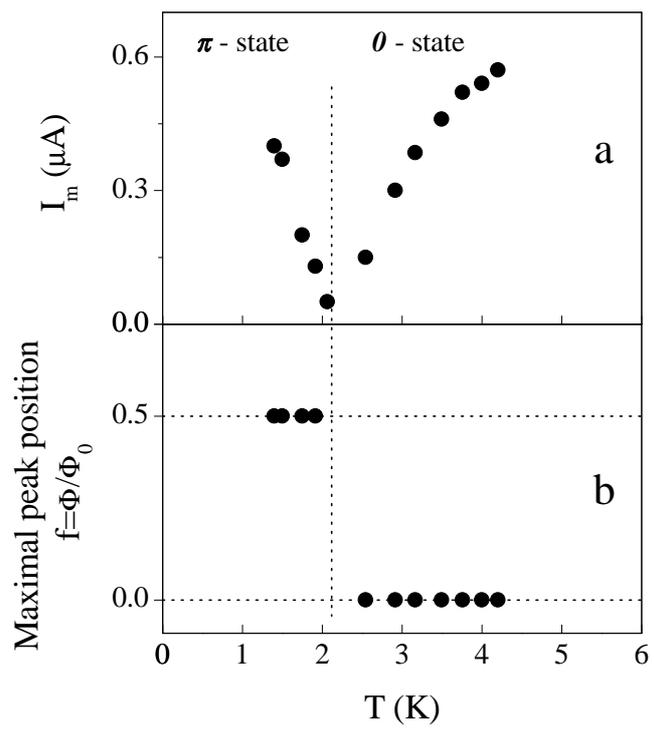
\end{document}